\documentclass[useAMS,usenatbib,usegraphicx]{mn2e}
\usepackage{color}

\title[PBHs from the EPA epoch]{Primordial Intermediate and Supermassive Black Hole formation during the electron-positron annihilation epoch}
\author[J. L. G. Sobrinho, Augusto P.]{J. L. G. Sobrinho$^{1}$\thanks{E-mail: josens@staff.uma.pt }, P. Augusto$^{2}$\thanks{E-mail: pedroaugusto@med.up.pt }\\
$^{1}$Faculdade de Ci\^{e}ncias Exatas e da Engenharia, Universidade da Madeira, Campus da Penteada, 9020-105 Funchal,  Portugal\\
$^{2}$Faculdade de Medicina da Universidade do Porto, Al. Prof. Hern\^ani Monteiro, 4200-319, Porto, Portugal}

\begin{document}

\date{Accepted 2024 April 4. Received 2024 April 1; in original form 2023 December 19}

\pagerange{\pageref{firstpage}--\pageref{lastpage}} \pubyear{2014}

\maketitle

\label{firstpage}

\begin{abstract}

Some of the Intermediate Mass Black Hole (IMBH) candidates observed at the center of galaxies or in globular clusters and some of the Supermassive Black Holes (SMBHs) seen at the center of many galaxies might be of primordial origin. Indeed,  Primordial Black Holes (PBHs) of such mass could have formed when the Universe was $\sim$1--10$^3$~s old, due to the collapse of density fluctuations. In particular, when the Universe was $\sim 1$~s in age, Electron-Positron Annihilation (EPA) took place. We explore the formation of intermediate mass and supermassive PBHs, taking into account the  effect of the EPA when the fluctuations have a running-tilt power-law spectrum: when these cross the $10^{-0.5}$--$10^{3.0}$~s Universe horizon they could produce $5\times 10^{3}$--$5\times 10^{8}M_{\odot}$ PBHs with a density  as high as $\sim 10^{10}$/Gpc$^3$. On average, this implies a population of about one thousand PBHs in the Local Group of Galaxies, with the nearest one at about 
250~kpc, just under half the distance to the Andromeda galaxy (M31).

\end{abstract}


\section{Introduction}
\label{sec:Introduction}

Primordial Black Holes (PBHs) may have formed in the early Universe due to the gravitational collapse of density fluctuations (with amplitude $\delta$ larger than a threshold $\delta_c$) with their masses, typically, given by the horizon mass, $M_{H}$, at the epoch $t_k$ of collapse (the horizon crossing time) when that fluctuation, which was disconnected from physical processes due to inflation, reenters the cosmological horizon \citep*[see][and references therein]{2016MNRAS.463.2348S}.
For a radiation-dominated Universe, in which the background sound speed is given by $c_{s,0}=1/\sqrt{3}$, the value of $\delta_c$ is a constant such that $0.4\leq \delta_c \leq 0.6$, with the true value being a function of the shape of the energy density perturbation. Here we consider $\delta_c\approx 0.50$, which corresponds to the typical Mexican-Hat perturbation profile \citep*[][]{2021PhRvD.103f3538M,2023arXiv230307980M}. 

In special circumstances, such as the QCD phase transition,  the value of $\delta_c$ may drop to as low as $\delta_c \approx 0.1$, due to the decrease on the background sound speed \citep[][]{2016MNRAS.463.2348S} which, in turn, could lead to an important population of PBHs
\citep[][]{2020MNRAS.496...60S}. Less dramatic reductions on the value of $\delta_c$ may also occur during cosmic annihilation epochs such as the  $e^{+}e^{-}$ annihilation \citep[e.g.][]{1999PhRvD..59l4014J,2020ARNPS..70..355C,2021PDU....3100755C,2023arXiv230307980M}. Although \citet{1997PhRvD..55.5871J} pointed to a decrease of $\sim 10$--20\% on the background sound speed value for a few Hubble times, more recently \citet{2021PDU....3100755C} mention a value around 5--10\% with \citet{2023arXiv230307980M} and \citet{2023PhRvD.108j1304A} reducing this to $\approx 5\%$-6\%. We consider the $e^{+}e^{-}$ annihilation (EPA) epoch  the period during which the sound speed stays bellow its background value.

Assuming that during the EPA the entropy is conserved (Universe evolving close to thermal equilibrium) the sound speed can be written as 
\citep{1999PhRvD..59d3517S}:
\begin{equation}
\label{sound_speed_square_general}
c_{s}^{2}(T)=\left(\frac{d\ln(s(T))}{d\ln(T)}\right)^{-1}
\end{equation}  
Here $s(T)$ is the entropy density \citep[e.g.][]{1990eaun.book.Kolb,2016Galax...4...78H}, given by
\begin{equation}
\label{entropy_density}
s(T)=\frac{2\pi^{2}}{45}g_{s}(T)T^3
\end{equation} 
where $g_{s}(T)$ accounts for the number of entropic degrees of freedom \citep{1990eaun.book.Kolb}:
\begin{equation}
\label{gs-freedom}
g_{s}(T)=\sum_{bosons}g_{i}\left(\frac{T_i}{T}\right)^{3} +\frac{7}{8}\sum_{fermions}g_{i}\left(\frac{T_i}{T}\right)^{3}
\end{equation} 
The factor $7/8$ takes into account the difference between Fermi and Bose
statistics and $T_i$ represents the possibility that the particle species $i$ may have an equilibrium distribution different from that of photons when the temperature is $T$ \citep{1990eaun.book.Kolb}.

When the temperature was $\approx 5$~MeV the Universe was mainly populated by relativistic photons, electrons, positrons and neutrinos, all of them in thermal equilibrium at the same temperature ($T_{i}/T=1$)  giving $g_{s}=10.75$ \citep{2008cosm.book.Weinberg}. When the temperature drops bellow $\approx 0.8$~MeV neutrinos decouple from the rest of the Universe since their (weak) interaction rate becomes lower than the expansion rate of the Universe \citep{2000ihep.book.Perkins,2002thas.book.Padmanabhan}.

When the temperature becomes lower than $0.511$~MeV (corresponding to the electron rest mass $m_e$) the production of new $e^{+}e^{-}$ pairs is severely suppressed, with the remaining ones annihilating. The energy released in this process reheats the photons but not the neutrinos since they are already decoupled. Thus, from this point on, photons and neutrinos evolve with different temperatures. At the end of the EPA epoch the temperature of the neutrinos $T_{\nu}$ relates to the temperature of  the photons $T_{\gamma}$ according to  \citep[e.g.][]{2003AnP...515..220S,2008cosm.book.Weinberg}:
\begin{equation}
\label{gs-assimptotic}
\frac{T_{\nu}}{T_{\gamma}}=\left(\frac{4}{11}\right)^{1/3}
\end{equation}
which holds until the present time, with $g_{s}\approx 3.91$ \citep[e.g.][]{1990eaun.book.Kolb,2003AnP...515..220S,2016Galax...4...78H}. During the EPA epoch the value of $T_{\nu}$ evolves according to \citep[][]{2008cosm.book.Weinberg}:
\begin{equation}
\label{function-alpha-sound}
T_{\nu}=T\left(\frac{4}{11}\right)^{1/3}\left(S\left(\frac{m_e}{T}\right)\right)^{1/3}
\end{equation}
where $S$ is a non-trivial function \citep[see][for details]{2008cosm.book.Weinberg}.

An  Intermediate Mass Black Hole (IMBH) is, by convention, a black hole with a larger mass than a Stellar mass Black Hole (SBH) but with a smaller mass when compared to a Supermassive Black Hole (SMBH): from $\sim100M_{\odot}$  to $10^6M_{\odot}$  \citep*[see, e.g.,][]{2020ARA&A..58..257G}. Given its location in cosmological time, it has been mentioned by others that the EPA epoch might have enhanced the formation of IMBHs and/or SMBHs \citep[e.g.][]{1997PhRvD..55.5871J,2021PDU....3100755C,2024PhR..1054....1C}.

It is now well-established that SMBHs reside in the centres of many galaxies \citep[e.g.][]{2013ARA&A..51..511K,2016ApJ...831..134V} including our own galaxy with a $4.297\times 10^{6}M_{\odot}$ SMBH \citep[][]{2023A&A...677L..10G}. Still in the Local Group of Galaxies (LG)  we  have SMBHs at the core of M31 \citep[$1.4\times 10^{8}M_{\odot}$;][]{2005ApJ...631..280B} and M32 \citep[$2.4\times 10^{6}M_{\odot}$;][]{2010MNRAS.401.1770V}.

So far, it has not yet been possible to demonstrate beyond reasonable doubt the existence of a single IMBH.
Nevertheless, a few strongest IMBH candidates have been identified
  \citep[][]{2020ARA&A..58..257G}. Out of these, we picked the eleven with a small mass uncertainty and show them in Table~\ref{tabela-IMBHs}. Their masses go from $2\times10^3$ M$_{\odot}$ to $9\times10^5$ M$_{\odot}$ and their distances from just 5~kpc to almost 1~Gpc.
Although several mechanisms concerning the formation of IMBHs have been extensively explored \citep[e.g.][]{2022ApJ...929L..22R}, we cannot rule out that some of them are primordial in origin.

\begin{table}
\caption[]{Intermediate Mass Black Hole candidates with a `known' mass (the error is less than one order of magnitude and there are no conflicting results in the literature), ordered by distance. For each candidate we show {\bf (1)} the  host name, {\bf (2)}  distance in Mpc, and {\bf (3)} its mass.
The data were retrieved from \citet{2020ARA&A..58..257G}. 
\label{tabela-IMBHs}
}
\center
\begin{tabular}{ccc}
\hline \hline
{\bf (1)} & {\bf (2)} & {\bf (3)} \\
 Host & Distance   & Mass  \\
  & (Mpc)   & ($\times 10^5 M_{\odot}$)
\\ \hline
47~Tuc		&	0.005		& 0.02\\
NGC 1904	&	0.013		& 0.03\\
NGC 5102	&	3.2				& 9\\
NGC 5206	&	3.5				& 6\\
UGC 6728	&	27				& 5\\
iPTF16fnl	&	67			& 3\\
ASASSN-14ae	&	200				& 3\\
WINGS J1348	&	265				& 5\\
PTF-09axc	&	536				& 5\\
PS1-10jh	&	822				& 7\\
PTF-09djl	&	900				& 7\\
\hline
\end{tabular}
\end{table}

The aim of this letter is to explore the  effect of the EPA epoch on PBH formation and how this could contribute to the population of IMBHs and SMBHs. The letter is organized as follows: in Section~\ref{sec:PBH formation during EPA} we evaluate the behavior of the threshold $\delta_c$ during the EPA epoch  and determine the number density of PBHs formed in a few specific situations. In Section~\ref{sec:discussion} we discuss the  results.


\begin{figure}
\centering
\includegraphics[width=84mm]{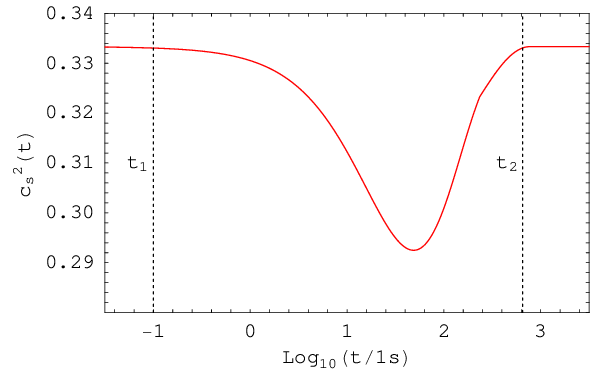}
\caption[]{\protect The sound speed $c_{s}^{2}(t)$ during the EPA epoch. Between the instants $t_{1}\approx 0.1\mathrm{~s}$ and $t_{2}\approx 650\mathrm{~s}$ the sound speed stays below  $c_{s,0}^{2}=1/3$, reaching a minimum of $c_{s,min}^{2}\approx 0.292$ at $t\approx  50\mathrm{~s}$.}
\label{epa-cs2-2024}
\end{figure}

\begin{figure}
\centering
\includegraphics[width=84mm]{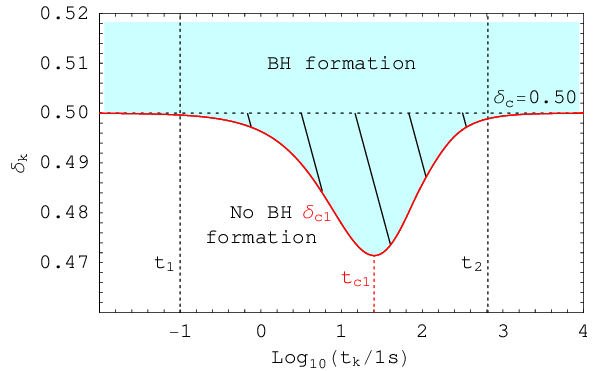}
\caption[]{ \protect 
The ($\log_{10}(t_{k}/\mathrm{1s}), \delta_{k}$) plane wherein the shadowed region there is collapse to a PBH.
The curve gives the limits  for the EPA epoch model when  $\delta_{c}=0.50$, with the vertical lines $t_{1}\approx 0.1\mathrm{~s}$ and $t_{2}\approx 650\mathrm{~s}$ defining the EPA epoch (cf. Figure~\ref{epa-cs2-2024}). For a given horizon crossing time, $t_{k}$, the dashed region represents our newly found window for PBH formation. The minimum of $\delta_{c1}$ is 0.471 at $t_{k}=t_{c1}\approx 25\mathrm{~s}$. 
\label{sobrinho2007_2024_EPA_crossover_total}}
\end{figure}

\section{PBH formation during the EPA epoch}
\label{sec:PBH formation during EPA}

Using equations~(\ref{sound_speed_square_general}), 
(\ref{entropy_density}), (\ref{gs-freedom}), 
and (\ref{function-alpha-sound})
 we determined the behavior of the sound speed during the EPA epoch (Figure~\ref{epa-cs2-2024}): $c_{s}^{2}(t)$ stays bellow its background value ($c_{s,0}=1/\sqrt{3}$) between the instants $t_1\approx 0.1$~s and $t_2\approx 650$~s with the minimum value $c_{s,min}^{2}\approx 0.292$ (which corresponds to a reduction of $\approx 6\%$) reached when $t\approx 50$~s.

The value of the threshold $\delta_{c}\approx 0.50$ changes slightly during the EPA when the horizon crossing time $t_k$  gets close enough to reach a new PBH formation threshold $\delta_{c1}<\delta_c$. In particular when $t_{k}=t_{c1}\approx 25\mathrm{~s}$ we have $\delta_{c1}\approx 0.471$ (see \citet[][]{2016MNRAS.463.2348S} for details on the method used to evaluate $\delta_{c1}$) which corresponds to the smallest value attained by $\delta_{c1}$ (Figure \ref{sobrinho2007_2024_EPA_crossover_total}).

For the density fluctuations we considered a running-tilt power-law spectrum \citep[cf. equation (4) of][]{2020MNRAS.496...60S}. In order to compute the corresponding spectral index, $n(k)$, we take the observational values of the parameters $n_0$, $n_1$ and $n_2$ measured at some pivot scale $k_c$ (Table~\ref{tabela-inicial}). In order to explore different scenarios we attribute values to the still unknown parameters $n_3$ and $n_4$ and assume that $n_i=0$ when $i\geq5$ \citep[see Section~3 of][]{2020MNRAS.496...60S}.
We then relate ($n_3,n_4$) to the more meaningful quantities ($n_{max},t_{k_{max}}$), $n_{max}$ being the maximum value attained by the spectral index $n(k)$ and $t_{k_{max}}$ the instant when  $n_{max}$ takes place \citep[cf. Equations (18) and (19) of][]{2020MNRAS.496...60S}.

\begin{table*}
\caption[]{Parameters used in this paper (some only implicitly). Note that most values are updated from the Table~\ref{tabela-IMBHs} of \citet[][]{2020MNRAS.496...60S}. References for the last column: 
[1]~\citet[][]{2020A&A...641A...6P};
[2]~\citet[][]{2022PTEP.2022h3C01W}.
\label{tabela-inicial}
}
\center
\begin{tabular}{cp{10cm}lc}
\hline
\hline
{\bf Parameter} & \multicolumn{1}{c}{\bf Description}  & {\bf Value}  & {\bf Reference} \\
\hline
$n_{0}$ & spectral index at the pivot scale ($k_c$) & 0.9647 & [1]\\\\
$n_{1}$ & running of the spectral index & 0.0011 & [1]\\\\
$n_{2}$ & running of the running of the spectral index  & 0.009 & [1]\\\\
$k_{c}$ & pivot scale & $0.05\mathrm{~Mpc}^{-1}$ 
& [1,2]\\\\
$\delta_{H}^{2}(k_{c})$ & amplitude of the density perturbation spectrum at the pivot scale  ($k_c$) & $2.0989\times10^{-9}$ & [2] \\\\
$\rho_{c}(t_{0})$ & critical density of the Universe at the current epoch (t$_0$)  & $8.531\times 10^{-27}\mathrm{~kgm}^{-3}$ & [2] \\\\
$\Omega_{CDM}$ & Cold Dark Matter density parameter  & $0.265$ & [2] \\
\hline
\end{tabular}
\end{table*}

\begin{figure}
\centering
\includegraphics[width=80mm]{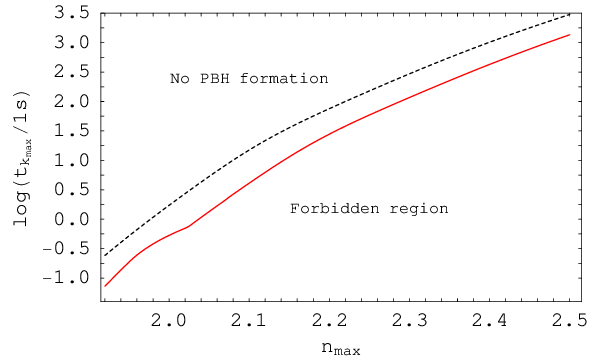} \\
\caption[]{The curves in the $(n_{max},\log(t_{k_{max}}/1\mathrm{~s}))$ plane indicating which parameter values lead to PBH formation during the EPA epoch. Below the solid curve, PBH formation is not allowed since it would violate the observational constraints \citep[cf.][]{2021RPPh...84k6902C}. Above the dashed curve, PBH formation is allowed although in negligible numbers (less than one PBH within the observable Universe). The region of interest, as regards PBH formation, is the one located between the two curves, with the most favourable situations on the solid curve: the number density of PBHs decreases as one moves from the solid curve towards the dashed one. See Figure~\ref{fig4-beta}.
\label{fig3-n3n4}}
\end{figure}

We found, numerically \cite[with][]{wolfram}, that the EPA epoch is well covered by considering $1.920\leq n_{max}\leq 2.500$ and $10^{-1.0}\mathrm{~s}\leq t_{k_{max}}\leq 10^{3.0}$~s. 
In Figure~\ref{fig3-n3n4} we  show the region on the $(n_{max},\log(t_{k_{max}}/1\mathrm{~s}))$ plane where PBH formation is possible, sitting between
i) the  `Forbidden region',  where the amount of PBHs would violate the observational constraints \citep[cf.][]{2021RPPh...84k6902C}; and ii) `No PBH formation', actually meaning that their numbers are negligible (less than one PBH within the entire observable Universe). For a given value of  $t_{k_{max}}$ the fraction of the Universe going into PBHs, $\beta (t_k)$ \citep[see][]{2020MNRAS.496...60S}, will be maximum if the corresponding value of $n_{max}$ is the one located over the solid curve in Figure~\ref{fig3-n3n4}.

\begin{figure}
\centering
\includegraphics[width=65mm]{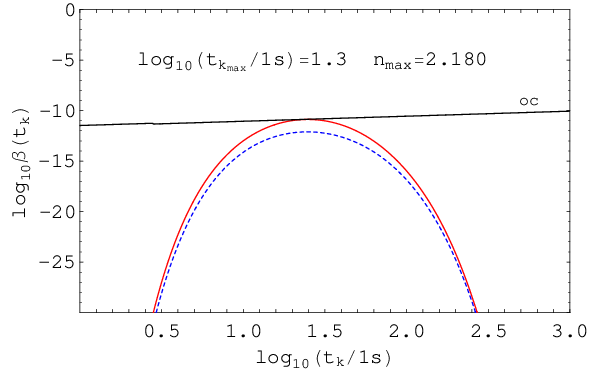} \\
\caption[]{\protect  The fraction of the Universe going into PBHs ($\beta(t_k)$) during the EPA epoch when $t_{k_{max}}=10^{1.3}\mathrm{~s}$ and $n_{max}=2.180$ (red solid line). The blue dashed line represents the curve $\beta(t_k)$ when the EPA effect is neglected. The curve labeled `oc' close to the top represents the observational constraints. This graph is the situation when the two $\beta(t_k)$ curves are farther from each other.
\label{fig4-beta}}
\end{figure}

In Figure~\ref{fig4-beta} we show the $\beta(t_k)$ curves for the case when $t_{k_{max}}=10^{1.3}\mathrm{~s}$ ($n_{max}=2.180$) and when one neglects the EPA contribution, which corresponds to the maximum difference between the two curves.

In Table~\ref{tabela-n-gpc3-epa} we show the PBH mass spectrum for a selection of cases (we pick up from the results of \citet[][]{2020MNRAS.496...60S} who cover up to $\sim 5\times10^2 M_{\odot}$) all of them located over the solid curve in Figure~\ref{fig3-n3n4}. 

\begin{table*}
\caption[]{\protect  The mass spectrum for a selection of specific situations which imply maximum PBH formation during the EPA epoch (on the solid line of Figure~\ref{fig3-n3n4}). {\bf (1:)} $t_{k_{max}}$, the instant when the spectral index attains its maximum value; {\bf (2:)} $n_{max}$, the maximum value attained by the spectral index; {\bf (3-8:)}  the number density of PBHs for each representative mass. 
\label{tabela-n-gpc3-epa}
}
\center
\begin{tabular}{cccccccc}
\hline \hline
{\bf (1)} & {\bf (2)} & {\bf (3)}  & {\bf (4)} & {\bf (5)} & {\bf (6)} & {\bf (7)} &  {\bf (8)}\\
$\log (t_{k_{max}}/1\mathrm{~s})$ & $n_{max}$ & \multicolumn{6}{c}{N/Gpc$^3$} \\
\cline{3-8}\\
 &   & $5\times10^{3}M_{\odot}$ &  $5\times10^{4}M_{\odot}$ & $5\times10^{5}M_{\odot}$ & $5\times10^{6}M_{\odot}$ &  $5\times10^{7}M_{\odot}$ &  $5\times10^{8}M_{\odot}$ \\
\hline
-0.5	& 1.970 & $8.9\times10^{7}$ 	& $2.6\times10^7$ 		& 0 					& 0  	& 0 & 0\\
	
0		& 2.036 & $1.6\times10^{7}$ 	& $1.8\times10^{10}$ 	& $3.0\times10^{6}$ 	& 0 	& 0 & 0\\ 
		
0.5  	& 2.087 & 0 					& $1.9\times10^{9}$ 	& $2.0\times10^{9}$ 	& 0  	& 0 & 0\\ 

1.0  	& 2.143 & 0 & $7.2\times10^{3}$ 	& $1.4\times10^{9}$ 	& $6.5\times10^{5}$ &  0 & 0\\ 

1.5  & 2.207 & 0 & 0 & $1.4\times10^{8}$ & $1.9\times10^{8}$ & 0 & 0\\ 

2.0  & 2.288 & 0 & 0 & $1.4\times10^{2}$ & $1.2\times10^{8}$ & $4.4\times10^{4}$ & 0\\ 

2.5  & 2.376 & 0 & 0 & 0 & $6.3\times10^{6}$ & $2.7\times10^{7}$ & 0\\ 

3.0  & 2.475 & 0 & 0 & 0 & 0 & $9.1\times10^{6}$ & $4.7\times10^{4}$\\ 
\hline
\end{tabular}
\end{table*}

\begin{figure}
\centering
\includegraphics[width=84mm]{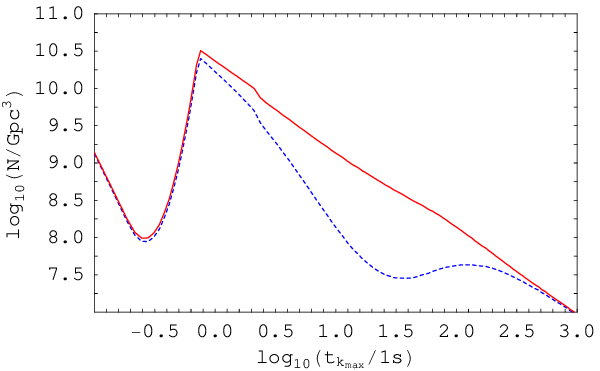}
\caption[]{\protect  The {\em maximum} number of PBHs per Gpc$^3$ that could form as a function of $t_{k_{max}}$ when the EPA is taken into account (red solid curve) and when the effect of the EPA is neglected (blue dashed curve). 
Note that the largest EPA contribution occurs for $t_{k_{max}}\simeq 20$~s (this corresponds to the case represented in Figure~\ref{fig4-beta}).}
\label{fig5-n-R-EPA}
\end{figure}

In Figure~\ref{fig5-n-R-EPA} we show the curves giving the maximum number of PBHs that could form as a function of $t_{k_{max}}$ when the EPA effect is taken into account and when it is not. 
Moving from left to right we have a region for which the two curves coincide (up to $t_{k_{max}}\simeq 1$~s) followed by a section where the contribution of the EPA clearly dominates, with the largest contribution from the EPA occurring  for $t_{k_{max}}\simeq 20$~s (cf.~Figure~\ref{fig4-beta}). When  $t_{k_{max}}\simeq 10^3$~s the two curves converge again.

\section{Discussion}
\label{sec:discussion}

Although no IMBH has been securely confirmed as of today, there is every reason to anticipate their existence, since many well-studied mechanisms might form them, including the primordial Universe
($\leq 10$~s old), which includes the EPA epoch. PBHs can form in the early Universe due to the  collapse of overdense regions provided that the amplitude $\delta$ of the density fluctuations is greater than some threshold $\delta_c$.  Although during the radiation-dominated Universe $\delta_c$ remains constant ($\delta_c\approx 0.50$) it might decrease if the Universe goes through a cosmological phase transition or an annihilation epoch such as the EPA,   favoring PBH formation. 
Indeed, during the EPA epoch, we get a result as low as
$\delta_c= 0.471$ ($\approx 6\%$ less than its background value). As a consequence, we have  discovered that, in the case of a running-tilt power-law spectrum, the EPA  increases the number density of PBHs  when $t_{k_{max}}$ (the instant when the maximum spectral index of the fluctuations takes place) occurs close to the EPA epoch (see Figure~\ref{fig5-n-R-EPA}).

\citet[][]{2021PDU....3100755C}  explored the effect of the EPA contribution to the number density of PBHs by considering different cases that always peak at SMBH masses ($\sim 10^6M_{\odot}$). Our results cover a much broader range, from IMBH to SMBH masses  ($\sim 10^3$-$10^8 M_{\odot}$).   

Indeed, as regards SMBHs we have obtained number densities of $\sim 10^{8}$/Gpc$^3$ (for  $t_{k_{max}}\simeq 30$-100~s and $\sim5\times 10^{6}M_{\odot}$), giving  two-to-three PBHs within the LG which is consistent with what we know, so far.
It is plausible to consider that PBHs within this mass range would grow up due to accretion \citep[e.g.][]{2021PDU....3100755C}. As for IMBH, their number density could reach $\sim 10^{10}$/Gpc$^3$ (for  $t_{k_{max}}\simeq 1~s$ and $\sim5\times 10^{4}M_{\odot}$), giving  $\sim 250$ PBHs within the LG with the nearest one at an average distance of $\sim0.25$~Mpc (assuming an homogeneous distribution), $\sim10^{9}$/Gpc$^3$ (for  $t_{k_{max}}\simeq 3~s$  and $\sim5\times 10^{4}M_{\odot}$-$5\times 10^{5}M_{\odot}$), giving  $\approx 60$ PBHs within the LG, with the nearest one at an average distance of $\sim0.5$~Mpc, or $\sim10^{9}$/Gpc$^3$ (for  $t_{k_{max}}\simeq 10~s$  and $\sim5\times 10^{5}M_{\odot}$), giving  $\approx 20$ PBHs within the LG, with the nearest one at an average distance of $\sim0.6$~Mpc. In all the three mentioned cases the nearest IMBH is closer than the Andromeda galaxy (M31).

Up to now we are aware of 11 IMBH candidates with a known mass (cf. Table~\ref{tabela-IMBHs}). Considering a region with radius $\approx 900$~Mpc (the distance to the farthest IMBH in Table~\ref{tabela-IMBHs}) we get an IMBH number density of $\approx 4/$~Gpc~$^3$ which is a value that falls far too short of those presented in Table~\ref{tabela-n-gpc3-epa}. However if we consider only the four IMBH candidates which are closest to us ($\leq 3.5$~Mpc; roughly the size of the LG) then we come up with a number density of $\sim 10^7 /$Gpc$^3$ which is more in accordance to our  results. 

The vast majority of known IMBH candidate masses lying on the $\sim10^{5}$--$10^{6}M_{\odot}$ mass range might
 reflect the difficulty on the detection of lighter IMBHs. In fact, at the present time, it is not easy to distinguish between an IMBH candidate and other types of sources when one considers masses bellow $\sim 10^4M_{\odot}$ although dynamical and accretion signatures point to a fraction of at least $50\%$ of galaxies with masses of $10^9$--$10^{10}M_{\odot}$ to host a $10^4$--$10^{6}M_{\odot}$ IMBH \citep[e.g.][]{2020ARA&A..58..257G}. 
We believe there are, at least, two types of observational bias at play here: i) the most obvious one is related to the weakness of the sources that lie much beyond the LG; ii) we are still looking for the best IMBH confirmation technique, since the current ones have results that are often in conflict with each other \citep[e.g.][]{2020ARA&A..58..257G}.



\section*{Data Availability}

All data used in this letter are available upon request from the authors.


\end{document}